\documentclass[trackchanges,twocolumn]{aastex701}

\usepackage{subcaption}

\begin{document}

\title{The Inner and Outer Shock Layers of Bow Shocks in Cataclysmic Variables}

\author[orcid=0000-0002-4005-5095]{Krystian I\l{}kiewicz}
\affiliation{Nicolaus Copernicus Astronomical Center, Polish Academy of Sciences, Bartycka 18, 00-716 Warsaw, Poland}
\email[show]{ilkiewicz@camk.edu.pl}

\author[0000-0002-1116-2553]{Christian Knigge}
\affiliation{School of Physics and Astronomy, University of Southampton, Highfield, Southampton SO17 1BJ, UK}
\email{C.Knigge@soton.ac.uk}

\author[0000-0001-5387-7189]{Simone Scaringi}
\affiliation{Centre for Extragalactic Astronomy, Department of Physics, Durham University, DH1 3LE, UK}
\affiliation{INAF-Osservatorio Astronomico di Capodimonte, Salita Moiariello 16, I-80131 Naples, Italy}
\email{simone.scaringi@durham.ac.uk}

\author[0000-0002-5870-0443]{Noel Castro Segura}
\affiliation{Astronomy and Astrophysics Group, Department of Physics, University of Warwick, Coventry CV4 7AL, UK}
\email{Noel.Castro-Segura@warwick.ac.uk}

\author[0000-0002-5761-2417]{Santiago del Palacio}
\affiliation{Department of Space, Earth and Environment, Chalmers University
of Technology, SE-412 96 Gothenburg, Sweden}
\email{santiago.delpalacio@chalmers.se}

\author[0000-0002-0146-3096]{Martina Veresvarska }
\affiliation{Institute of Space Sciences (ICE, CSIC), Campus UAB, Carrer de Can Magrans s/n, 08193, Barcellona, Spain}
\affiliation{Institut d’Estudis Espacials de Catalunya (IEEC), Esteve Terradas 1, RDIT Building, 08860, Castelldefels, Spain }
\email{mveresvarska@ice.csic.es}

\begin{abstract}
Bow shocks around cataclysmic variables (CVs) have traditionally been identified with a single bright optical arc. This feature has been interpreted as the bow shock formed by the interaction between a sustained outflow and the interstellar medium (ISM). We show that this interpretation is incomplete. Generic wind--ISM interaction theory predicts a two-shock configuration consisting of an inner terminal wind (reverse) shock and an outer forward shock, separated by a hot, low-density shocked wind cavity. Using archival ultraviolet, optical, and infrared imaging of the nova-like systems BZ~Cam and V341~Ara, and the polar 1RXS~J052832.5+283824, we find that the nebulae around all three systems exhibit this layered structure. In each case, the previously identified bow shock bright in H$\alpha$ and [O\,{\sc iii}] corresponds to a compact inner arc, while additional emission components reveal a more extended morphology. Specifically, each system shows an outer arc detected in mid-infrared images, and the region between the optical and infrared arcs is filled with faint H$\alpha$ emission and, where available, far ultraviolet emission. We identify this infrared arc, reported here for the first time in these systems, as the sweep-up boundary of the forward shock, while the bright inner optical arc corresponds to the terminal wind shock rather than the forward shock as previously assumed. These results reveal that the true extent and layered structure of bow shocks around CVs only become apparent when observations extend beyond the optical band.

\end{abstract}


\section{Introduction}\label{sec:intro}

Cataclysmic variables (CVs) are close interacting binaries in which a late-type donor transfers mass to a white dwarf (WD). The accretion geometry is set largely by the WD magnetic field: weakly magnetic systems form an accretion disk, intermediate polars host truncated disks with magnetically controlled accretion curtains, and polars channel the flow directly along field lines onto the WD poles. Some CVs, and in particular high–mass-transfer nova-like systems, also launch sustained outflows. These winds are most clearly revealed through P~Cygni profiles of emission lines in the ultraviolet and optical range, indicating mass loss with characteristic velocities of order a few $10^{2}$--$10^{3}\,$km\,s$^{-1}$ \citep[e.g.,][]{1992ApJ...393..217H,2001A&A...376.1031G}.

When a source with a sustained outflow moves through the interstellar medium (ISM), the wind–ISM interaction can produce a bow shock. Persistent, spatially resolved bow shocks have been identified around only a handful of CVs to date \citep[e.g.,][]{1987A&A...181..373K,2018PASP..130i4201B,2019MNRAS.486.2631H,2025arXiv251103587B,nature_bowshocks}, but those systems provide direct constraints on the long-term momentum and energy input from compact binaries into their local environments \citep{2021MNRAS.501.1951C,nature_bowshocks}. 

BZ~Cam is a nova-like with the best studied bow shock among CVs. The BZ~Cam bow shock is often portrayed as dominated by a single bright optical arc seen in H$\alpha$ and [O\,{\sc iii}], a feature that has been interpreted as the forward shock or H\,{\sc ii} front produced by the wind--ISM interaction \citep{1992ApJ...393..217H}. Subsequent work, however, has shown that the bow shock morphology and excitation are more complex than this single‑layer picture implies. Namely, deeper imaging and kinematic studies reveal extended, asymmetric structure and spatially varying ionization that are not easily reconciled with a single, steady forward shock \citep{2001A&A...376.1031G}. Moreover, theoretically idealized wind–ISM interactions produce a two‑shock geometry consisting of an inner reverse/terminal shock that decelerates and heats the fast outflow and an outer forward shock that sweeps up and compresses the ambient medium, with a large hot shocked‑wind cavity between them \citep{1992ApJ...393..217H}. These regions differ by few orders of magnitude in density and temperature, so their observational signatures can be very different. As a result, mapping these theoretical zones to observational tracers is non-trivial and multi-wavelength imaging is required to disentangle terminal‑shock, forward‑shock, and photoionized components.

In this work we test whether CV bow shocks display the generic two-shock wind--ISM structure. We focus on the three best-characterized systems: nova-like systems BZ~Cam and V341~Ara, and the diskless polar 1RXS~J052832.5+283824 (RXJ0528+2838). We combine archival ultraviolet and infrared imaging with published narrow-band H$\alpha$ and [O\,{\sc iii}] maps.  By comparing the spatial distributions of emission across these wavelength regimes, we assess how different observational tracers relate to the underlying shock structure and evaluate whether the commonly adopted single-layer interpretation of CV bow shocks provides a complete description. This approach allows us to place the observed nebulae in the broader context of wind-driven bow shocks.

\section{Data selection}\label{sec:data}

We searched archival images for bow shocks associated with BZ~Cam, V341~Ara, and RXJ0528+2838 using survey data accessible through the CDS/Virtual Observatory \citep{2000A&AS..143...33B}. For each target, we inspected all available survey observations and selected the datasets offering new insights into the structure of the bow shocks.

The bow shock of BZ~Cam is detected in the far-ultraviolet (FUV; 1350–1750 \AA), but not at the longer wavelengths covered by the near-ultraviolet detectors (NUV; 1750–2800 \AA) of the \textit{Galaxy Evolution Explorer} \citep[\textit{GALEX;}][]{2007ApJS..173..682M}. This ultraviolet bow shock was previously reported in \textit{Swift} observations by \citet{2014MNRAS.445..869Z}, but was not analyzed in detail. We note that the UV ring-like feature reported by \citet{2014MNRAS.445..869Z} in the UVOT image is located at a larger off-axis angle than the CV, consistent with known UVOT ghost artifacts. The corresponding UVOT quality map produced with {\tt uvotflagqual} flags this region, and no counterpart is detected in our \textit{GALEX} data. Together, these facts indicate that the ring is an instrumental artifact. For V341~Ara, no far-ultraviolet observations are available, and, as in BZ~Cam, no bow shock features are detected in the near-ultraviolet images. For RXJ0528+2838, no \textit{GALEX} observations exist.

In the infrared the BZ~Cam bow shock is visible in Unblurred Coadds of the \textit{WISE}  \citep[unWISE;][]{2014AJ....147..108L} longer wavelength W2 images (3.96--5.34~\(\mu\)m), but not in the shorter wavelength W1 band images (2.75--3.87~\(\mu\)m). The bow shocks of RXJ0528+2838 and V341~Ara are also present in W2, though their detailed morphology is harder to analyze because of the relatively low spatial resolution of WISE. Specifically, in V341~Ara the bow shock is less pronounced due to strong crowding in the field, while in RXJ0528+2838 the nebula is partially blended with the nearby bright infrared source 2MASS~J05283260+2837556.

To increase the visibility of the bow shocks in RXJ0528+2838 and V341~Ara, we scaled the W1 images to match the mean stellar brightness in W2 and subtracted W1 from W2, thereby removing most of the stellar contamination. We note that because the W2 point spread function is larger than that in W1, this procedure introduces residual artifacts in the form of rings at the locations of bright stars.

The optical narrow-band images of BZ~Cam are taken from \citet{2018PASP..130i4201B}, for V341~Ara from \citet{2021MNRAS.501.1951C}, and for RXJ0528+2838 from \citet{nature_bowshocks}.

\section{Results: layered bow-shock morphology}\label{sec:results}

\subsection{BZ~Cam: the predicted two-shock stratification}\label{sec:bzcam}

\begin{figure*}[ht!]
\centering
\includegraphics[width=0.75\textwidth]{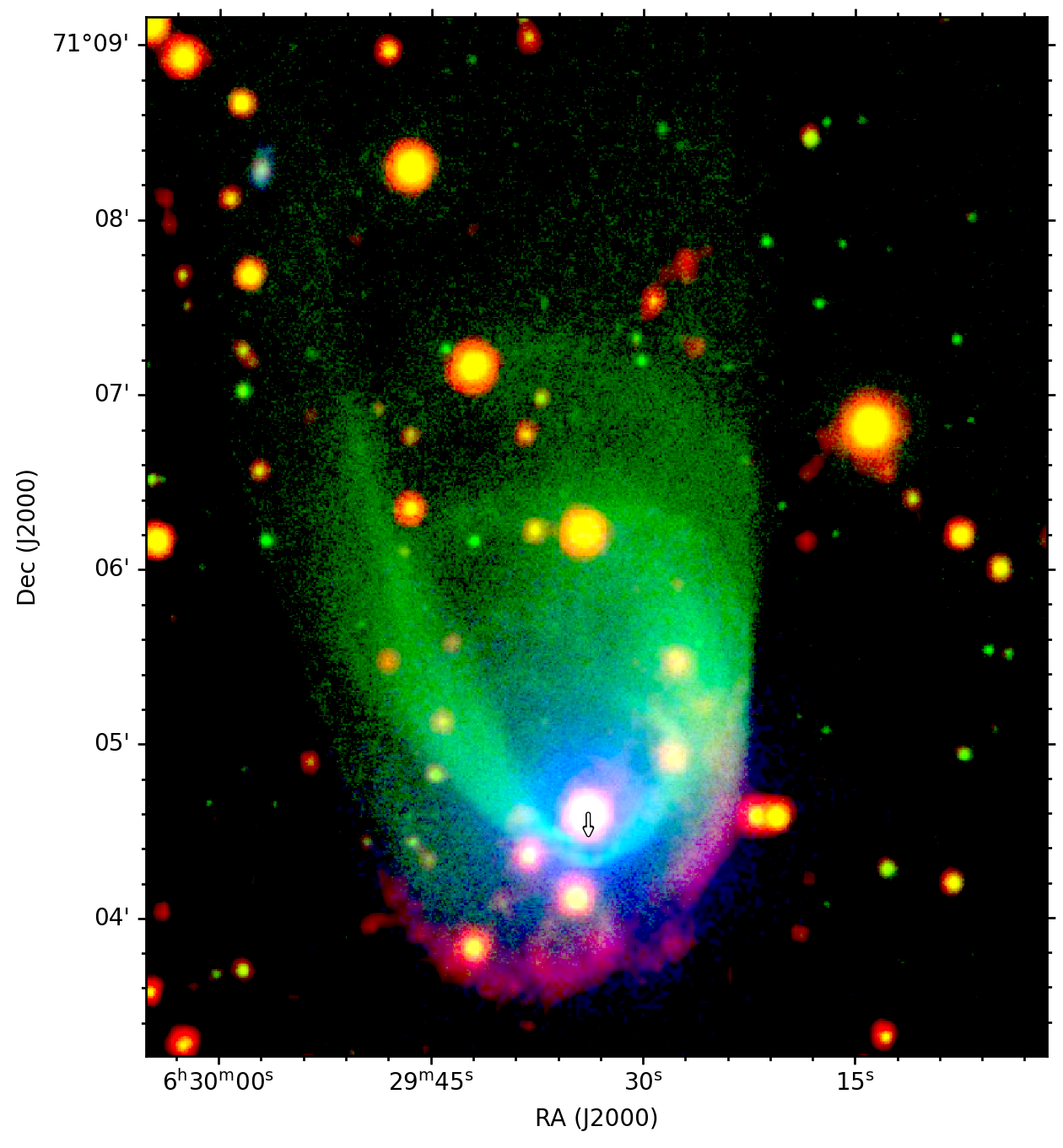}
\caption{False-color composite image of BZ~Cam: \textit{GALEX} FUV (blue), H$\alpha$ (green; \citealt{2018PASP..130i4201B}), and unWISE W2 (red). The morphology is stratified: a compact inner optical bow shock, an extended FUV/faint-H$\alpha$ interior, and an outermost W2 shell. The arrow indicates the Gaia proper-motion vector \citep{2023A&A...674A...1G} corrected for Galactic rotation. The corrected proper motion is $(\mu_{\alpha},\mu_{\delta}) = (-0.69\pm0.02,\,-21.61\pm0.07)\ \mathrm{mas\ yr^{-1}}$. The corresponding tangential, radial, and total space velocities are $V_t = 38.4\pm0.4$ km~s$^{-1}$, $V_r = -69\pm10$ km~s$^{-1}$, and $V = 79.3\pm8.8$ km~s$^{-1}$. North is up, east is left.}
\label{fig:bzcam_uvIRopt}
\end{figure*}

Figure~\ref{fig:bzcam_uvIRopt} compares the far-ultraviolet (FUV), H$\alpha$, and W2 morphologies of BZ~Cam bow shock. The emission is clearly stratified into three spatially distinct components.

\emph{(i) Inner optical bow shock:} The compact bow-shaped arc bright in H$\alpha$ and [O\,{\sc iii}]~$\lambda5007$ lies closest to the CV. This structure corresponds to the bow shock discussed in previous optical studies and has historically been treated as the primary manifestation of the wind--ISM interaction \citep[e.g.,][]{1992ApJ...393..217H,2001A&A...376.1031G}.

\emph{(ii) Intermediate faint layer:} Beyond the bright optical arc, a broader emission region bright in far-ultraviolet is visible which seems to be accompanied by a fainter H$\alpha$ component.  This extended component is consistent with emission from hot, low-density gas and/or strong far-ultraviolet lines commonly associated with shocked winds (e.g., C\,{\sc iv}, He\,{\sc ii}, N\,{\sc v}), although spectroscopy is required to determine the dominant contributors. This more extended H$\alpha$ emission is clearly visible in previous publications, though it has not been explicitly discussed by the authors \citep[e.g.,][]{2001A&A...376.1031G}.

\emph{(iii) Outer dust bow:} the most extended structure detected in W2 defines the outer edge of a larger bow shock. The W2 bow shock extends to $\sim55''$, much larger than the $\sim15''$ bright optical arc. The appearance in W2 and not W1 indicates cool dust emission concentrated near the swept-up boundary. Although the infrared bow axis remains aligned with the proper motion of BZ~Cam, the infrared morphology appears more asymmetric than that of the inner optical bow shock, with the infrared bow apex shifted farther east of the CV proper motion.  

\begin{figure}[ht!]
\centering
\includegraphics[width=0.49\textwidth]{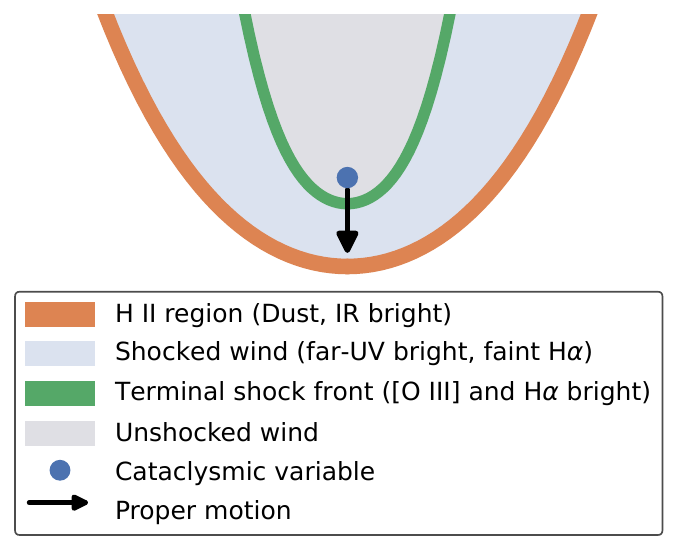}
\caption{Schematic mapping between observed tracers and physical zones in the two-shock interpretation of CV bow shocks. The outermost layer (orange) represents the swept-up H\,{\sc ii}/forward-shock boundary traced by infrared-emitting dust. The inner bow shock (green) represents the terminal/reverse shock front, which is bright in [O\,{\sc iii}] and H$\alpha$. The region between the two bow shocks (blue) is the shocked-wind cavity expected to be bright in the far-ultraviolet and only faintly in H$\alpha$. The inner region (gray) indicates unshocked wind with no clear observational signatures. The arrow shows the CV proper-motion direction.}
\label{fig:model_schematic}
\end{figure}

This ordering, with a high ionization optical bow shock inside, far-ultraviolet emission filling the cavity, and an outer infrared rim, is consistent with theoretical models of wind–ISM interactions, provided that the bright inner optical arc is associated with the terminal-wind (reverse) shock, or with photoionized wind material immediately upstream of the terminal shock. In this interpretation, the outer infrared shell traces the forward shock propagating into the ISM and the associated compressed, swept up dust layer (i.e., the H\,{\sc ii} region/forward shock). A schematic mapping between the observed tracers and the physical zones of the BZ~Cam bow shock is shown in Fig.~\ref{fig:model_schematic}. The observed asymmetry of the infrared bow shock likely reflects inhomogeneities in the local ISM, with the bow-shock surface expanding preferentially toward regions of lower external ram pressure \citep[e.g.][]{2019MNRAS.484.4760B}. In this interpretation, the eastward displacement of the infrared apex relative to the direction of proper motion implies a lower ISM density toward the southeast than toward the southwest, although asymmetry in the outflow cannot be ruled out. The differing asymmetries observed in the infrared and optical bow shocks may arise because infrared emission traces dust near the outer boundary of the interaction region, where coupling to the large-scale ISM is strongest, whereas H$\alpha$ and [O III] emission originates closer to the CV, where the flow is partially shielded from ISM density variations by the forward shock.

The infrared bow shock exhibits a somewhat clumpy morphology. Several of the brightest knots correspond to sources in the WISE All-Sky Source Catalog (WISEA), each with a unique identifier (e.g., WISEA~J062924.50+710430.7). Using the catalog photometry \citep{2014yCat.2328....0C}, we identified eleven such clumps with no clear optical counterparts. Four have measurements in the W1, W2, and W3 bands and only upper limits in W4, while the remaining seven knots are detected only in W1 and W2.

For the four clumps with three-band detections, we fitted simple blackbody models to their spectral energy distributions, obtaining dust temperatures of $\sim$500–600~K (see Appendix~\ref{appendix:wise}). Because the SEDs are constrained by only three photometric points, the blackbody fits should be interpreted with caution and are intended primarily to demonstrate the presence of warm dust rather than to characterize the detailed nature of the mid-infrared emission. These temperatures are nevertheless similar to those inferred at the apices of bow shocks around massive stars \citep{2014A&A...567A..21S}, where dust is heated by a combination of radiative and shock-related processes. In the case of BZ~Cam, the dust is unlikely to be heated predominantly by radiation from the CV given its relatively low luminosity and the large bow shock standoff distance, and is more likely heated by irradiation from shock-generated UV photons. However, more detailed observations will be required before firm conclusions can be drawn regarding the dominant dust heating mechanism.

We also extracted \textit{WISE} multi-epoch photometry for the clumps in the W1 and W2 bands using the procedures of \citet{2020MNRAS.493.2271H}.\footnote{\url{https://github.com/HC-Hwang/wise_light_curves}} None of the clumps shows statistically significant variability, although the photometric uncertainties are large.

\subsection{RXJ0528+2838 and V341~Ara: the same stratification across accretion regimes}\label{sec:others}

\begin{figure*}[ht!]
\centering
\includegraphics[width=0.49\textwidth]{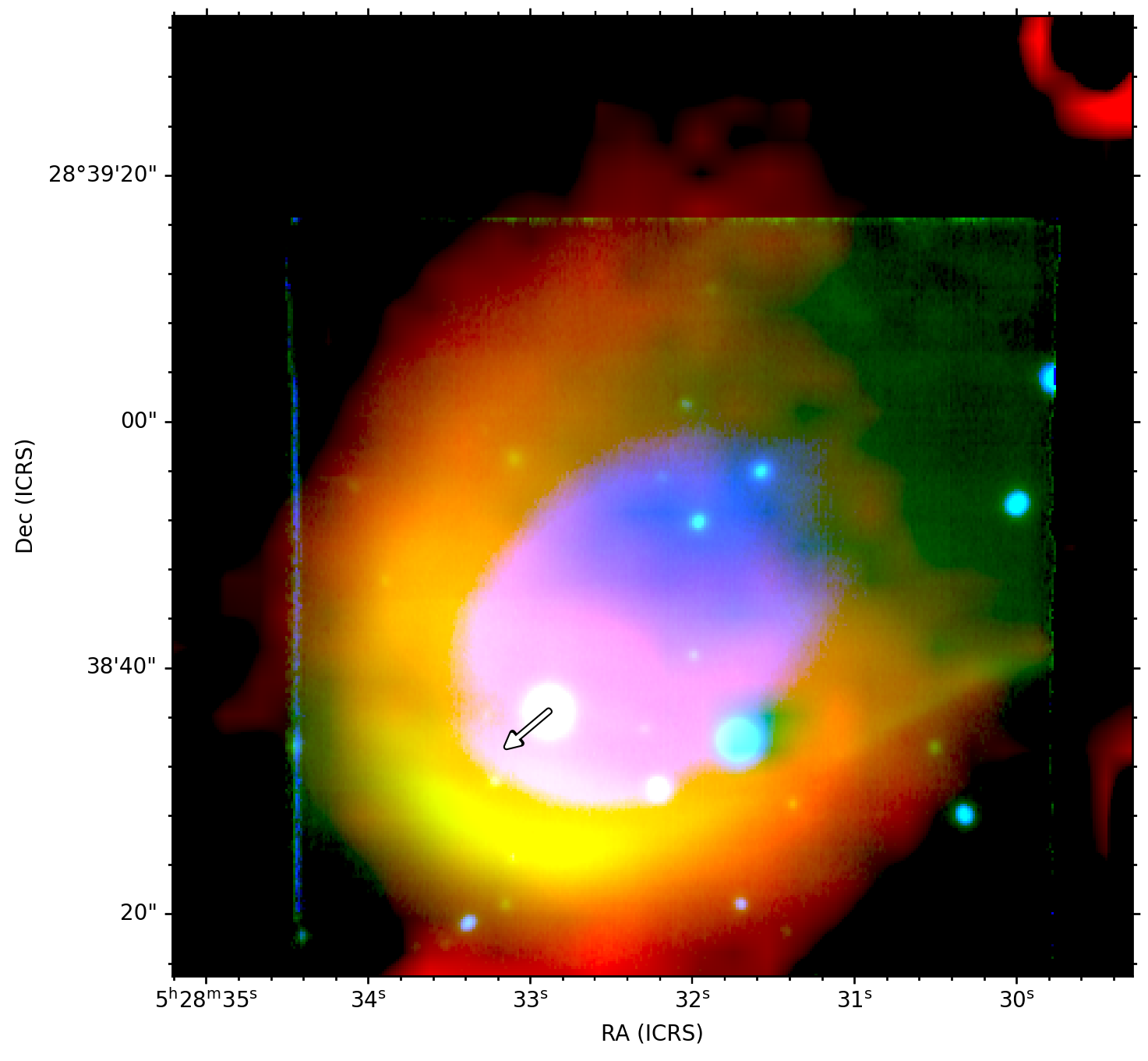}
\includegraphics[width=0.49\textwidth]{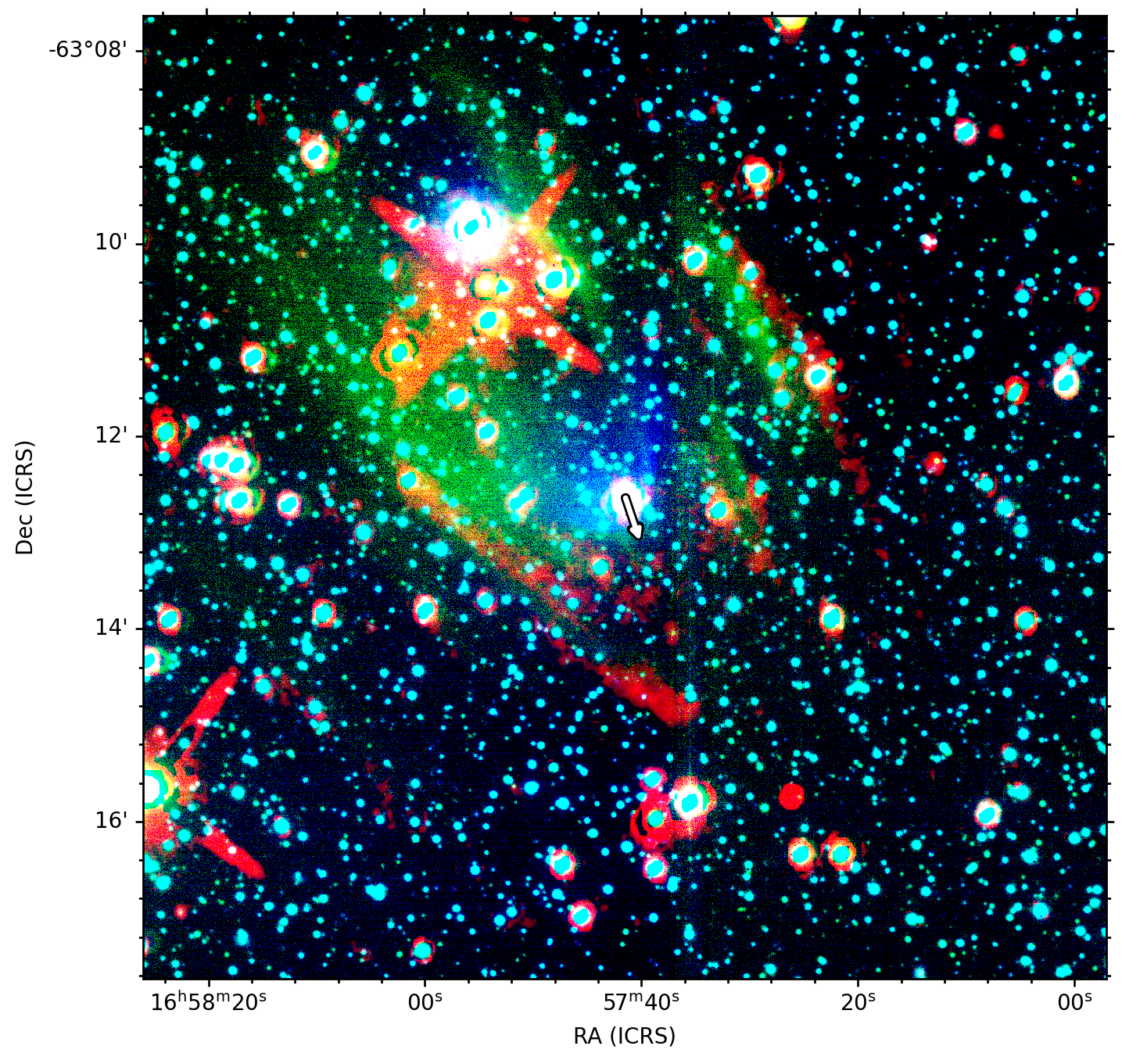}
\caption{False-color images of RXJ0528+2838 (top) and V341~Ara (bottom). Blue: [O\,{\sc iii}]~$\lambda5007$; green: H$\alpha$; red: W2--W1 difference (to suppress stellar crowding; see text). Optical narrow-band images are from \citet{nature_bowshocks} (RXJ0528+2838) and \citet{2021MNRAS.501.1951C} (V341~Ara). In both systems, the infrared emission marks the outermost bow shock layer. The arrows indicate the Gaia proper-motion vectors \citep{2023A&A...674A...1G} corrected for Galactic rotation. For RXJ0528+2838, the corrected proper motion is $(\mu_{\alpha},\mu_{\delta}) = (39.31\pm0.06,\,-29.1\pm0.2)\ \mathrm{mas\ yr^{-1}}$, corresponding to $V_t = 52.2\pm 0.9$ km~s$^{-1}$, $V_r = 117.4\pm 5.0$ km~s$^{-1}$, and a total space velocity $V = 128.5\pm 4.6$ km~s$^{-1}$. For V341~Ara, the corrected proper motion is $(\mu_{\alpha},\mu_{\delta}) = (-44.89\pm0.03,\,-60.2\pm0.1)\ \mathrm{mas\ yr^{-1}}$, with $V_t = 55.5\pm0.4$ km~s$^{-1}$, $V_r = 47.1\pm2.0$ km~s$^{-1}$, and $V = 72.7\pm1.3$ km~s$^{-1}$. North is up, east is left.}
\label{fig:1RXandV341_RGB}
\end{figure*}

In contrast to BZ~Cam, RXJ0528+2838 and V341~Ara lack far-ultraviolet imaging, and their infrared observations provide more limited constraints on the bow shock structure. In RXJ0528+2838 the nebula is compact and only marginally resolved, while in V341~Ara strong stellar crowding reduces the contrast of the infrared emission. As a result, we do not recover the same level of morphological detail as in BZ~Cam. Nevertheless, both RXJ0528+2838 and V341~Ara exhibit clear infrared bow shock emission in the unWISE W2 data (Fig.~\ref{fig:1RXandV341_RGB}), which traces cool dust at the outermost extent of their nebulae.

In RXJ0528+2838, the W2 emission forms a well defined parabolic arc that follows the outer boundary of the optical bow shock. Although the small angular size of the nebula ($\sim16.8''$) limits the amount of detail that can be recovered at the WISE resolution ($\sim6''$), the infrared arc is aligned with the system proper motion and reproduces the overall shape of the H$\alpha$ emission. The [O\,{\sc iii}] bow shock is confined to a smaller region interior to the W2 rim.

V341~Ara exhibits an analogous configuration on a larger angular scale. The W2 emission appears as an incomplete arc along the outer edge of the extended H$\alpha$ nebula previously discussed as a possible remnant nova shell \citep{2021MNRAS.501.1951C}. As in RXJ0528+2838, the infrared arc coincides with the outer boundary of the faint H$\alpha$ emission, while the [O\,{\sc iii}] emission highlights more internal, higher ionization structure. This morphology with outer infrared rim suggests that the larger H$\alpha$ structure is not a nova shell, but instead traces a shocked wind cavity similar to that seen in BZ~Cam. However, we note that a nova shell interpretation cannot be excluded by our observations, and that an abundance analysis, particularly of nitrogen in the shell, would be required to exclude or confirm a classical nova origin for the nebula.

In both systems, the infrared emission consistently marks the most extended component of the bow shock, whereas the [O\,{\sc iii}] bright optical emission is confined to a smaller bow shaped region closer to the central system. This spatial ordering mirrors the layered morphology identified in BZ~Cam and suggests that a similar stratification is present in RXJ0528+2838 and V341~Ara (Fig.~\ref{fig:model_schematic}), even though the absence of far-ultraviolet data prevents a direct detection of the shocked wind cavity.

\section{Discussion and Conclusions}\label{sec:conclusions}

We have used archival ultraviolet, optical, and infrared imaging to test whether bow shocks around cataclysmic variables exhibit the two-shock structure predicted by wind--ISM interaction theory. The combined multi-wavelength data for BZ~Cam, V341~Ara, and RXJ0528+2838 reveal a consistent, layered morphology that naturally maps onto the expected terminal-shock, shocked-wind, and forward-shock zones.

All three systems show an outer infrared arc detected in unWISE W2, reported here for the first time in each object. This infrared-bright rim marks the most extended component of the bow shock and traces the compressed, swept-up material at or near the forward shock. Its appearance in W2 but not W1 indicates emission from relatively cool dust concentrated in the outer boundary of the shocked region.

In BZ~Cam the region interior to the infrared rim is filled with far-ultraviolet emission and faint H$\alpha$, consistent with a hot, low-density shocked-wind cavity between the terminal and forward shocks. Although far-ultraviolet data are not available for V341~Ara or RXJ0528+2838, their morphologies show the same intermediate layer in the form of a faint, extended H$\alpha$ envelope.

The compact H$\alpha$+[O\,{\sc iii}] arcs in all systems correspond to the inner bow shock. This structure is the same feature that has traditionally been identified as \emph{the} bow shock in previous studies, but our results suggest it corresponds to the terminal/reverse shock front rather than the forward shock as previously assumed.

Taken together, these results show that CV bow shocks are fundamentally multi-layered structures: an inner terminal shock front producing the bright optical arc, a hot shocked-wind cavity that can emit strongly in the far-ultraviolet, and an outer forward-shock/H\,{\sc ii} boundary traced by infrared-emitting dust. The presence of this same layered structure in both disk-bearing nova-likes and in a diskless polar suggests that the observed morphology is governed primarily by the physics of the wind--ISM interaction, independent of the specific outflow-launching mechanism.

Future spatially resolved UV and infrared spectroscopy will be essential for determining the physical conditions within each layer and for constraining the long-term mass-loss histories of accreting white dwarfs.

\begin{acknowledgments}
This work was supported by the Polish National Science Centre (NCN) grant 2024/55/D/ST9/01713.
SS acknowledges support by the Science and Technology Facilities Council (STFC) grant ST/X001075/1.

This research made use of APLpy, an open-source plotting package for Python \citep{2012ascl.soft08017R}. This work has made use of data from the European Space Agency (ESA) mission
{\it Gaia} (\url{https://www.cosmos.esa.int/gaia}), processed by the {\it Gaia}
Data Processing and Analysis Consortium (DPAC,
\url{https://www.cosmos.esa.int/web/gaia/dpac/consortium}). Funding for the DPAC
has been provided by national institutions, in particular the institutions
participating in the {\it Gaia} Multilateral Agreement.

\end{acknowledgments}

\section*{Data Availability}
This work is based exclusively on archival data. No new observations were obtained by the authors. Some of the data presented in this article were obtained from the Mikulski Archive for Space Telescopes (MAST) at the Space Telescope Science Institute. The specific observations analyzed can be accessed via \dataset[doi:10.17909/cxk9-jt29]{https://doi.org/10.17909/cxk9-jt29}.

\facilities{WISE, GALEX, ESO(MUSE)}
\software{aplpy \citep{2012ascl.soft08017R}}

\bibliography{literature}{}
\bibliographystyle{aasjournalv7}

\appendix

\section{WISE knots in the BZ~Cam outer shell}\label{appendix:wise}

The W2 arc of BZ~Cam contains multiple compact clumps coincident with WISE All-Sky Source Catalog entries (WISEA identifiers). Using catalog photometry \citep{2014yCat.2328....0C}, we identified eleven such clumps lacking obvious optical counterparts. Four have W1--W3 measurements and only upper limits in W4, and the remaining seven are detected only in W1 and W2. For the four three-band clumps we fitted simple blackbodies by sampling the blackbody parameter space with a Markov Chain Monte Carlo method, obtaining characteristic temperatures of $\sim$500--600~K (Appendix Figure~\ref{fig:wise_sed_lc}). Multi-epoch W1/W2 photometry following \citet{2020MNRAS.493.2271H} shows no significant variability within the uncertainties.

\begin{figure*}[ht!]
\centering

\begin{subfigure}[t]{0.48\textwidth}
  \centering
  \includegraphics[width=\linewidth]{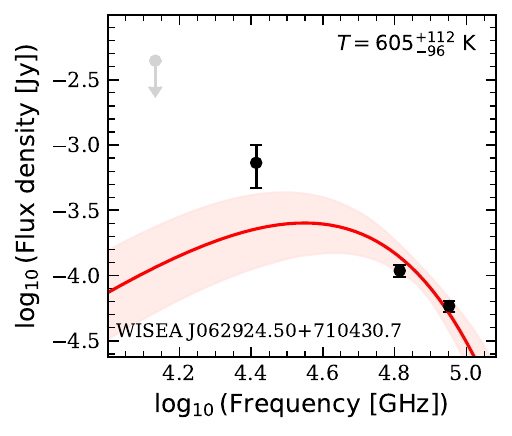}
\end{subfigure}\hfill
\begin{subfigure}[t]{0.48\textwidth}
  \centering
  \includegraphics[width=\linewidth]{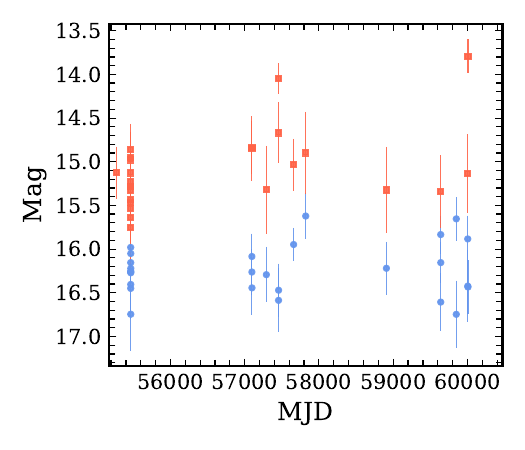}
\end{subfigure}

\vspace{1ex}

\begin{subfigure}[t]{0.48\textwidth}
  \centering
  \includegraphics[width=\linewidth]{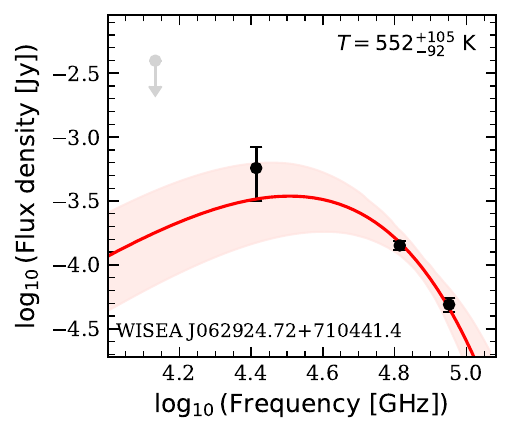}
\end{subfigure}\hfill
\begin{subfigure}[t]{0.48\textwidth}
  \centering
  \includegraphics[width=\linewidth]{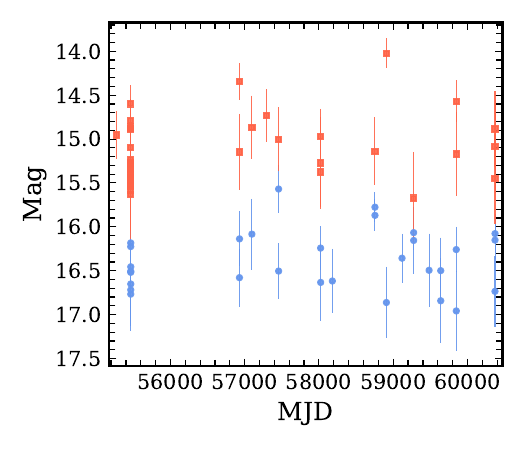}
\end{subfigure}

\caption{Spectral energy distributions (left) and W1/W2 light curves (right) for representative WISEA clumps on the BZ~Cam outer shell. Red curves show blackbody fits; shaded regions indicate fit uncertainties.}
\label{fig:wise_sed_lc}
\end{figure*}

\begin{figure*}[ht!]
\ContinuedFloat
\centering

\begin{subfigure}[t]{0.48\textwidth}
  \centering
  \includegraphics[width=\linewidth]{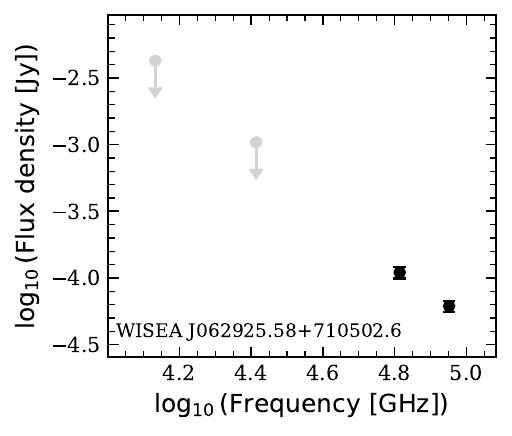}
\end{subfigure}\hfill
\begin{subfigure}[t]{0.48\textwidth}
  \centering
  \includegraphics[width=\linewidth]{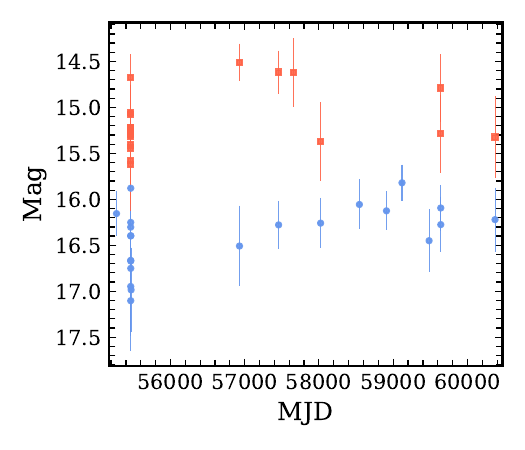}
\end{subfigure}

\vspace{1ex}

\begin{subfigure}[t]{0.48\textwidth}
  \centering
  \includegraphics[width=\linewidth]{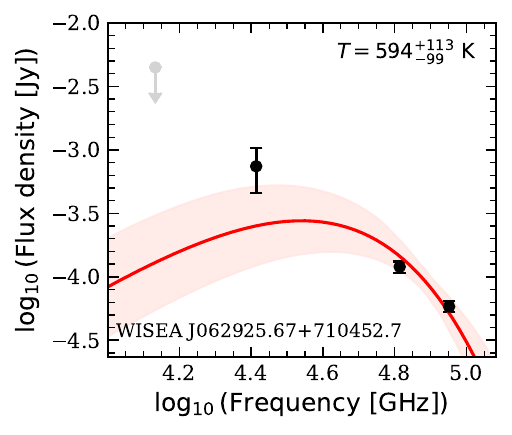}
\end{subfigure}\hfill
\begin{subfigure}[t]{0.48\textwidth}
  \centering
  \includegraphics[width=\linewidth]{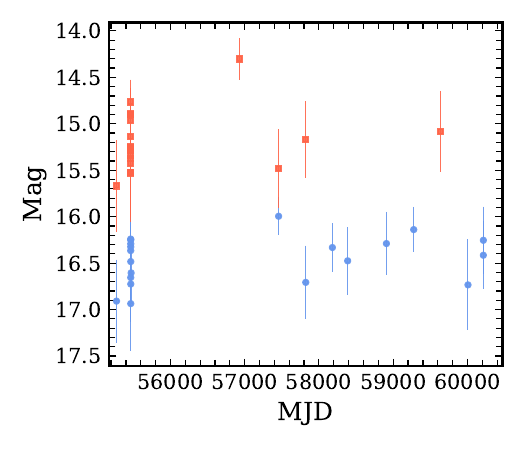}
\end{subfigure}

\vspace{1ex}

\begin{subfigure}[t]{0.48\textwidth}
  \centering
  \includegraphics[width=\linewidth]{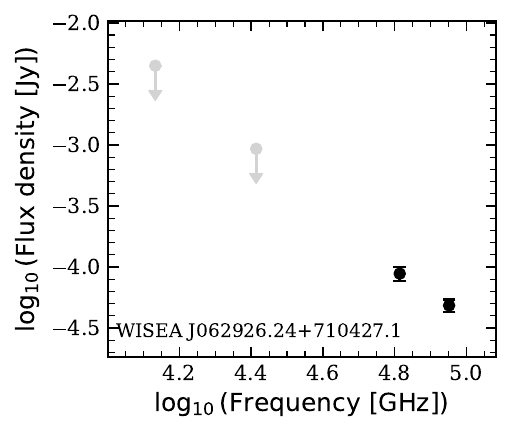}
\end{subfigure}\hfill
\begin{subfigure}[t]{0.48\textwidth}
  \centering
  \includegraphics[width=\linewidth]{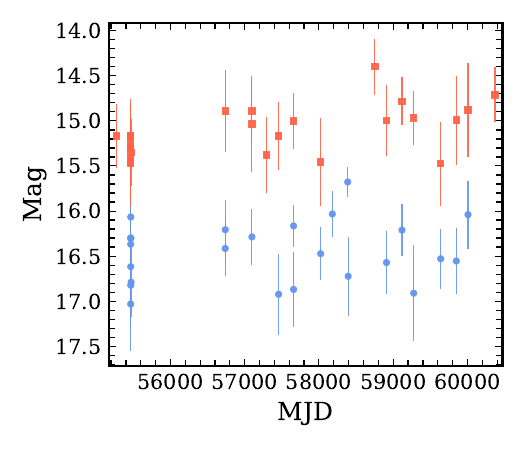}
\end{subfigure}

\caption{(Continued.)}
\end{figure*}

\begin{figure*}[ht!]
\ContinuedFloat
\centering

\begin{subfigure}[t]{0.48\textwidth}
  \centering
  \includegraphics[width=\linewidth]{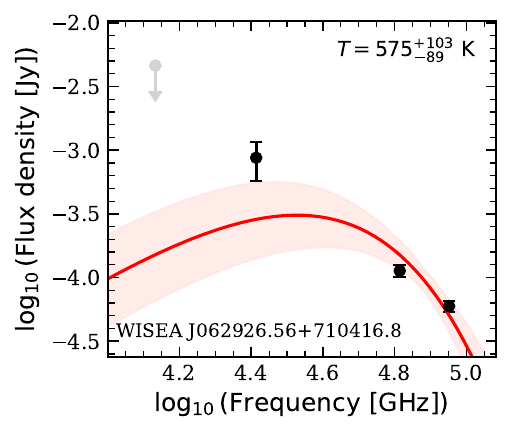}
\end{subfigure}\hfill
\begin{subfigure}[t]{0.48\textwidth}
  \centering
  \includegraphics[width=\linewidth]{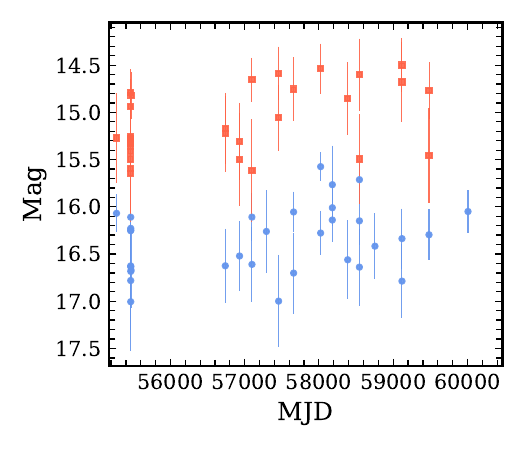}
\end{subfigure}

\vspace{1ex}

\begin{subfigure}[t]{0.48\textwidth}
  \centering
  \includegraphics[width=\linewidth]{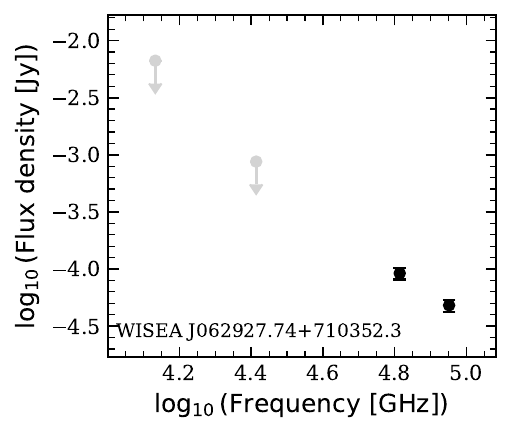}
\end{subfigure}\hfill
\begin{subfigure}[t]{0.48\textwidth}
  \centering
  \includegraphics[width=\linewidth]{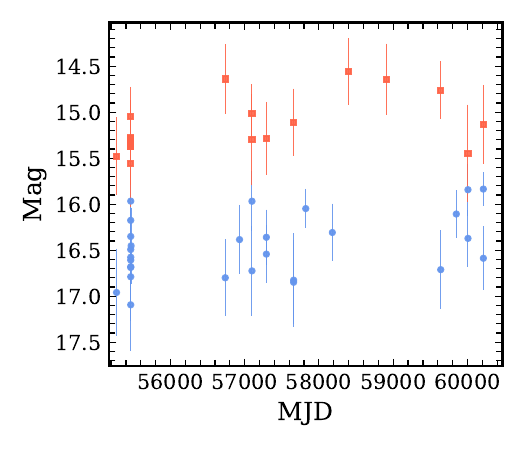}
\end{subfigure}

\vspace{1ex}

\begin{subfigure}[t]{0.48\textwidth}
  \centering
  \includegraphics[width=\linewidth]{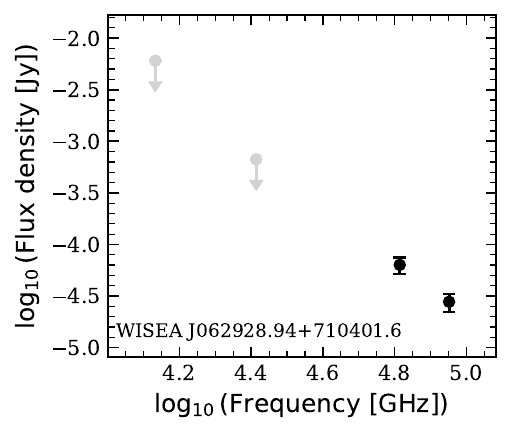}
\end{subfigure}\hfill
\begin{subfigure}[t]{0.48\textwidth}
  \centering
  \includegraphics[width=\linewidth]{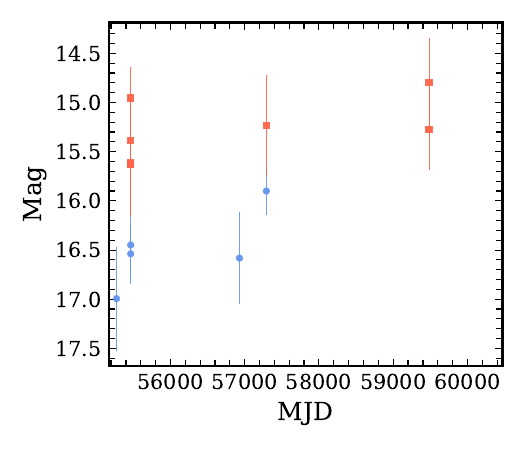}
\end{subfigure}

\caption{(Continued.)}
\end{figure*}

\begin{figure*}[ht!]
\ContinuedFloat
\centering

\begin{subfigure}[t]{0.48\textwidth}
  \centering
  \includegraphics[width=\linewidth]{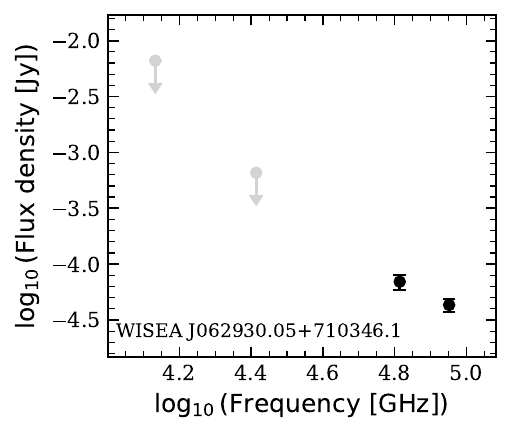}
\end{subfigure}\hfill
\begin{subfigure}[t]{0.48\textwidth}
  \centering
  \includegraphics[width=\linewidth]{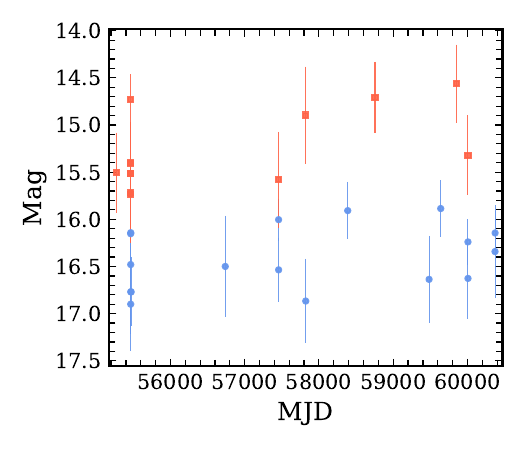}
\end{subfigure}

\vspace{1ex}

\begin{subfigure}[t]{0.48\textwidth}
  \centering
  \includegraphics[width=\linewidth]{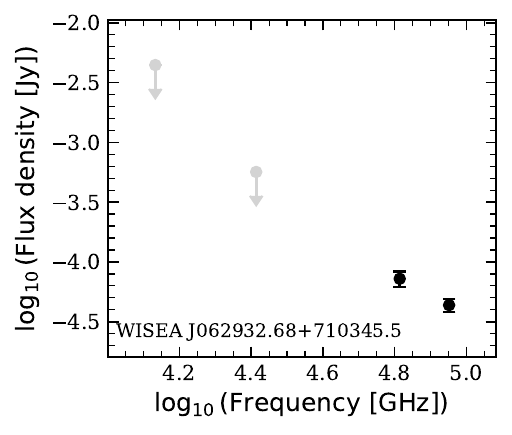}
\end{subfigure}\hfill
\begin{subfigure}[t]{0.48\textwidth}
  \centering
  \includegraphics[width=\linewidth]{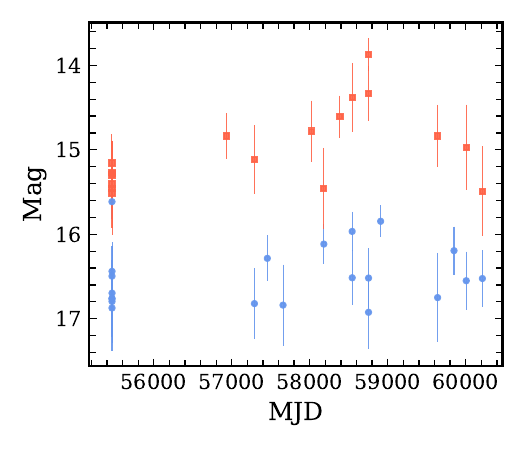}
\end{subfigure}

\vspace{1ex}

\begin{subfigure}[t]{0.48\textwidth}
  \centering
  \includegraphics[width=\linewidth]{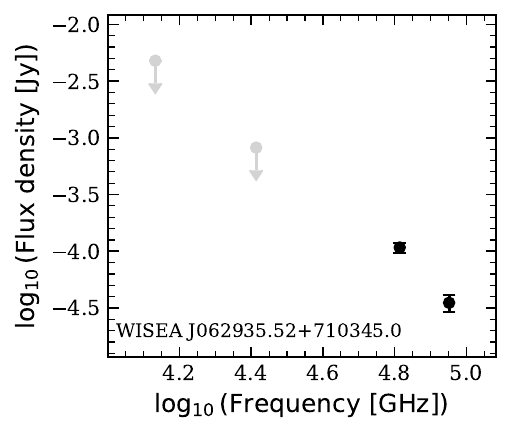}
\end{subfigure}\hfill
\begin{subfigure}[t]{0.48\textwidth}
  \centering
  \includegraphics[width=\linewidth]{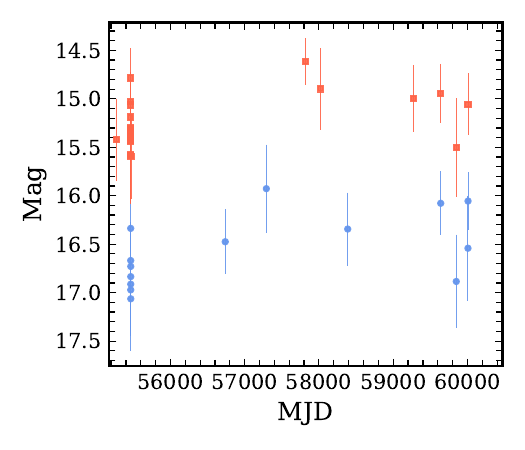}
\end{subfigure}

\caption{(Continued.)}
\end{figure*}

\end{document}